\documentclass[preprint,superscriptaddress,showpacs,keywords,preprintnumbers,amsmath,amssymb,aps,prb]{revtex4}
\usepackage{graphicx}
\usepackage{dcolumn}
\usepackage{bm}
\usepackage{color}
\begin{document}

\title{Minima of the fluctuations of the order parameter of seismicity worldwide}

\author{N. V. Sarlis}\email{nsarlis@phys.uoa.gr}
\affiliation{Department of Solid State Physics and Solid Earth Physics
Institute, Faculty of Physics, School of Science, National and Kapodistrian University of Athens,
Panepistimiopolis, Zografos 157 84, Athens, Greece}
\author{S.-R. G. Christopoulos}
\affiliation{Department of Solid State Physics and Solid Earth Physics
Institute, Faculty of Physics, School of Science, National and Kapodistrian University of Athens,
Panepistimiopolis, Zografos 157 84, Athens, Greece}
\author{E. S. Skordas}
\affiliation{Department of Solid State Physics and Solid Earth Physics
Institute, Faculty of Physics, School of Science, National and Kapodistrian University of Athens,
Panepistimiopolis, Zografos 157 84, Athens, Greece}

\begin{abstract}
It has been shown [Phys. Rev. E {\bf 84}, 022101 (2011); Chaos {\bf 22}, 023123 (2012)] that earthquakes of 
magnitude $M$ greater or equal to 7 are globally correlated. Such correlations were 
identified by studying the variance  $\kappa_1$ of natural time which has been proposed 
as an order parameter for seismicity. Here, we study the fluctuations of this order parameter using the Global Centroid 
Moment Tensor catalog for a magnitude threshold $M_{thres}=5.0$ and focus on its behavior before major earthquakes. Natural time analysis
reveals that distinct minima of the fluctuations of the order parameter of seismicity appear {within} almost five and a half months{-on average-} before
all major earthquakes of magnitude larger than 8.4. This phenomenon corroborates 
the finding [Proc. Natl. Acad. Sci. U.S.A. {\bf 110}, 13734 (2013)] 
that similar minima of the seismicity order parameter fluctuations  
had preceded all major shallow earthquakes in Japan.  
Moreover, on the basis of these minima a statistically significant
binary prediction method for earthquakes of magnitude  larger  than 8.4
with hit rate 100\% and false alarm rate 6.67\%  is suggested.

\end{abstract}

\maketitle

{\bf Earthquake (EQ) time series manifest complexity in many aspects showing complex correlations between space, time and magnitude $M$. 
Recently, it has been found that there exist correlations between the successive EQs of magnitude 7.0 or larger 
in a global scale. This suggests that the whole of the solid Earth crust should be considered as a single complex system. Adopting 
this view, we study here these correlations focusing on  the prediction of the time of occurrence of the strongest  (magnitude class 9) EQs in the globe. We make use
of a methodology recently applied  to the Japanese region that is based on natural time analysis which allows the introduction 
of an order parameter for seismicity. The study of the fluctuations of this order parameter has revealed distinct minima one to
three  months before the occurrence of all major EQs around Japan.  This kind of analysis is now generalized   for the period 1 January 1976 to 1 October 2014 considering all EQs with  magnitude greater than or equal to 5.0 in the Global
Centroid Moment Tensor catalog.
Considering a sliding natural time   window of length equal to the number of EQs that on average occur within a few months,
we show that similar minima can be also identified for the global seismicity. In particular, it is shown that all magnitude class 9 EQs, i.e, with magnitude greater than 8.4, have been preceded by such minima with an average lead time of five and a half months. 
This lead time almost coincides with the maximum lead time observed for the low frequency precursory electric signals termed Seismic Electric 
Signals (SES) activities, thus supporting the recent finding that the initiation of an SES activity marks the time when 
the system enters a critical stage. The properties
of these minima are investigated and a simple prediction method based on these minima is suggested. {Furthermore, the present results
strengthen the view that the solid Earth crust behaves as a single complex system.}  }

\section{Introduction}
\label{sec1}

Earthquakes (EQs) exhibit complex correlations in space, time and magnitude $M$ (e.g. see Refs. \onlinecite{SOR00,COR04,DAV05,SAI06,HOL06,EICH07,LEN08,LIP09,TEL09,TEL10B,BOT10,NEWTSA,LEN11,SAR11,SARCHRIS12A}) which are inevitably interrelated with the fundamental power laws of seismicity. Principal among the latter is the Gutenberg-Richter law for the 
 distribution of earthquake magnitudes
 which states that the
(cumulative) number of earthquakes with magnitude greater than $M$
occurring in a specified area and time is given by
\begin{equation}
N(\geq M)\sim 10^{a-bM} \label{eq10}
\end{equation}
where $b$ is a constant, which varies only slightly from region to
region  being generally in the range $0.8\leq b \leq 1.2$ (see
Ref. \onlinecite{RUN03} and references therein) and $a$ the logarithm of the earthquakes with magnitude greater than zero\cite{SHC04} measuring the intensity of regional seismicity\cite{TUR02} (note that Eq.(\ref{eq10}) holds both {\em regionally} and {\em
globally}). It is also well known that EQs are clustered in time and the Omori (power) law describes the temporal decay of aftershock
activity\cite{UTS61} (see also
Ref. \onlinecite{SHC04}):
\begin{equation}
r(t,M)=\frac{1}{\tau_0 \left[1+t/c(M) \right]^p},
\end{equation}
where $r(t,M)$ {denotes} the rate of occurrence of aftershocks with
magnitudes greater than $M$ per day, $t$ is the time that has
elapsed since the mainshock and $\tau_0$ and $c(M)$ are
characteristic times. Note that  $p$ is close to unity for large
earthquakes (e.g., see Ref. \onlinecite{SAI07}). 
There are also  power-laws referring to the distribution\cite{SOR91} $\sim 1/L^2$
of fault lengths $L$, the fractal structure of fault
networks\cite{DAV90} as well as the universal law for the
distribution of waiting times and seismic rates derived\cite{BAK02} from the analysis of space-time windows.
Considering seismicity as a (non-equilibrium) critical phenomenon, 
the observed\cite{TUR97} EQ power laws point to
 the proximity of the system to a critical point.\cite{HOL06} Adopting this view, 
 an order parameter
for seismicity labeled $\kappa_1$ has been introduced \citep{NAT05C} in the frame of
the analysis in a new time domain termed natural time $\chi$ (see
Section \ref{sec2}) which has been found to reveal novel dynamical
features hidden in the time series of complex systems.
\citep{SPRINGER}

The study of the order parameter $\kappa_1$ in global seismicity has revealed\cite{SAR11,SARCHRIS12A} that there exist correlations between successive EQ magnitudes when $M\geq 7.0$. These correlations have been identified by comparing the statistics of the experimentally observed $\kappa_1$ with those obtained when destroying by shuffling the order of occurrence of these EQs. The present paper focuses on the global fluctuations of $\kappa_1$ in view of the recent finding\cite{PNAS13} that in Japan such fluctuations exhibit characteristic minima before the occurrence of EQs of magnitude $M\geq 7.6$. In short, the distribution of $\kappa_1$ for the number of EQs that on average occur within a few months has been studied\cite{PNAS13} by means of the ratio $\beta$ (called\cite{NEWEPL} variability, see Section \ref{sec2})  of the  standard deviation over the mean of the corresponding $\kappa_1$ values and $\beta$ minima have been identified before the stronger EQs. 

Here, we analyze the Global Centroid Moment Tensor catalog, described in Section \ref{sec3}, in order to examine 
the existence of similar $\beta$ minima. The results, which are presented in Section \ref{sec4}, show that 
such $\beta$ minima do exist in global seismicity and precede magnitude class 9 EQs. These results are further discussed in Section \ref{sec5} and the corresponding conclusions are summarized in Section \ref{sec6}.

\section{Natural time analysis. Background}\label{sec2}
Natural time analysis has been shown\cite{ABE05} to {be optimal for enhancing the signals in time-frequency space}. 
For a time series comprising $N$ events, e.g. EQs, the natural time for the occurrence of the ${k-\rm{th}}$ event 
of energy $Q_k$ is defined\citep{NAT01,NAT02,NAT02A} by $\chi_k=k/N$. We then study the evolution of 
the pair $(\chi_k,p_k)$ where 
\begin{equation} \label{eq1}
p_k=Q_k/\sum_{n=1}^{N}Q_n
\end{equation}
is the normalized energy and construct the quantity $\kappa_1$ which is the variance of $\chi$ weighted by $p_k$  
\begin{equation}\label{k1}
\kappa_1=\sum_{k=1}^N p_k \chi_k^2 -\left( \sum_{k=1}^N p_k \chi_k \right)^2{\equiv \langle \chi^2 \rangle - \langle \chi \rangle ^2.}
\end{equation}
In the case of EQs, the quantity $Q_k$ of Eq.(\ref{eq1}) is the radiated seismic energy which is proportional\cite{KAN78} to the seismic scalar moment $M_0$ and can be estimated by means of the usual relation\cite{KAN78} 
\begin{equation}\label{kan}
Q_k \propto 10^{1.5 M_k},
\end{equation}
where $M_k$ is the moment magnitude of the ${k-\rm{th}}$ EQ.\cite{NAT05C,NAT09B,NEWEPL,VAR11,EPL12,EPL12A,NHESS,RAM13,TECTO12,FLO14} The quantity $\kappa_1$ has been proposed\cite{NAT05C} 
as an order parameter for seismicity since it changes abruptly to zero upon the occurrence of a strong event and exhibits\cite{NAT05C,NAT06B,SAR11,NAT11A,SARCHRIS12A} a behavior similar to that of the order parameter for various equilibrium and 
non-equilibrium critical systems. 

As mentioned in Section \ref{sec1}, the fluctuations of this order parameter versus conventional time can be studied\cite{NEWEPL,VAR11,EPL12,EPL12A,PNAS13,PNAS15}  by means of its variability $\beta$  within  excerpts of the EQ catalog ending just before a target EQ. The calculated $\beta$ value is assigned to this target EQ and hence to
 its (conventional) occurrence time. 
Specifically,\cite{PNAS15} let us consider an EQ catalog $\{ M_p\}, p=1,2,\ldots, L$ and its excerpts, labeled by $j=0,1,\ldots,L-W$, comprising  $W$ consecutive 
EQs, e.g., $\{ M_{j+m}\}, m=1,2,\ldots, W$. {For such an excerpt, we form its subexcerpts consisting of the $n^{th}$ to $(n+5)^{th}$ EQs, $(n=1,2, \ldots W-5)$ (since at least $l=6$ EQs are needed\cite{NAT05C} for obtaining reliable $\kappa_1$) and compute
 $\kappa_1$ through Eq.(\ref{k1}) for each of them by assigning $\chi_k=k/6$ and $p_k=10^{1.5M_{j+n+k-1}}/\sum_{k'=1}^{6}10^{1.5M_{j+n+k'-1}}$, $k=1,2,\ldots 6$, to the $k^{th}$ member of the subexcerpt.
 This procedure leads to $(W-5)$ values of $\kappa_1$. We iterate the same process for new subexcerpts comprising $l=7$ members 
 (giving rise to $(W-6)$ values of $\kappa_1$ ), 8 members \ldots and 
finally $W$ members. Then, we compute the average $\mu(\kappa_1)$ and the standard deviation $\sigma(\kappa_1)$ of the thus-obtained $(W-5)+(W-6)+\ldots +1[=(W-4)(W-5)/2]$ $\kappa_1$ values. The quantity $\beta_{W} (j)\equiv\sigma(\kappa_1) / \mu(\kappa_1)$ is  the variability of {the order parameter}
$\kappa_1$  for this excerpt of length $W$ labeled by $j$.  This}
$\beta_{W}(j)$  value   is assigned to the $j+W+1$-th EQ in the catalog, the
target EQ  that took place at (conventional) time $t_{j+W+1}$. Hence, for the $\beta_W$ value of a target EQ only its
past EQs are used in the calculation. The time evolution of the
$\beta_W$ value is then pursued by sliding the excerpt $W$ 
through the EQ catalog, i.e., by considering the pairs $(t_{j+W+1},\beta_W(j)), j=0,1, \ldots, L-W$, for examples see Fig. \ref{f2a}.

\section{The Data Analyzed} \label{sec3}
The data analyzed  come from the Global Centroid Moment Tensor  (CMT) Project\cite{DZI81,EKS12} that covers global seismicity since 1 January 1976. For EQs that took place before  2011, the 1976 to 2010 CMT catalog was used, whereas for EQs since 1 January 2011 to 1 October 2014 the monthly CMT catalogs have been employed (all these catalogs are publicly available from \url{http: \\www.globalcmt.org/CMTfiles.html}). The CMT catalog provides the scalar seismic moment $M_0$ for each listed event and thus this value was directly used as $Q_k$ in Eq.(\ref{eq1}) for the calculation of the $\kappa_1$ values.   In accordance with a previous study,\cite{SAR11}  we considered all EQs of magnitude greater than or equal to $M=5.0$. This resulted in 38,006 EQs, whose epicenters are shown in Fig. 1 of Ref. \onlinecite{Chaos15}, during the concerned period of 38 years and 9 months (1 January 1976 to 1 October 2014). Hence, we have a monthly rate of approximately 80 EQs.  Table \ref{tab2} summarizes all the magnitude class 9 EQs in CMT.

\section{Results} \label{sec4}
The purpose of the present study is, as mentioned in Section \ref{sec1},  to investigate whether there exist variability minima in the fluctuations of the order parameter $\kappa_1$ of global seismicity similar to those observed\cite{PNAS13,JGR14} in Japan. For this 
purpose, we examine $\beta_W$ when using as $W$ the number of events that on average occur within a few months.
The reasoning behind this selection is that Seismic Electric Signals (SES), which are low frequency ($\leq 1$Hz)
transient changes of the electric field of the Earth (the study of which has been motivated by solid state physics aspects on point defects, see Ref. \onlinecite{VAR08438} and references therein) that have
been found  to precede EQs, exhibit a lead time within this range of a few months. The latter has been verified by numerous observations of SES
activities in Greece\cite{VAR91,VAR93,VAR96B,NAT08,SPRINGER} which  agree with
later observations in {Japan \citep{UYE00,UYE02,UYE09,ORI12,XU013} and China\citep{HUA11B}}, as
well as in California (see \citet{BER91,FRA90} where magnetic
field variations similar to those associated with the SES
activities in Greece\cite{NAT09V} have been reported) and Mexico (see
\citet{RAM11} and references therein), which also showed that within a few months after the SES activity observation a strong main shock takes place. 
Thus, this time scale of a few months provides an estimate of the time in which the system enter{s} a critical stage before a strong EQ. 

Along these lines,  $\beta_{100}$, $\beta_{160}$, $\beta_{240}$, $\beta_{320}$ and $\beta_{400}$ have been plot in Fig.2 of Ref. \onlinecite{Chaos15} versus the conventional time. 
It is observed that as $W$ decreases the corresponding $\beta_{W}$ fluctuates rapidly becoming more difficult to interpret. {Indeed, when $W_2> W_1$, the quantity $\beta_{W_2}$ seems to smoothen the results obtained for $\beta_{W_1}$. This is so because the $\kappa_1$ values used for the calculation of various $\beta_{W_1}$s are included in the ensemble of $\kappa_1$ values from which $\beta_{W_2}$  is obtained (see Section \ref{sec2}). Thus, one might say that when increasing the window length from $W_1$ to $W_2$ apparently a  (non-linear) smoothing is performed. Along these lines, we start our analysis} from Fig.2(e) of Ref. \onlinecite{Chaos15}, we observe two clear precursory minima of $\beta_{400}$ with $\beta_{400}<0.26$: one before the 2004 Sumatra-Andaman  $M$9 EQ and another one before the 2007 Sumatra, Indonesia  $M$8.5 EQ (see Table \ref{tab2})  (cf. another minimum also appears  {\em but after} the 2012 Indian Ocean $M$8.6 EQ{, see the brown arrows in Fig.2(e) of Ref. \onlinecite{Chaos15} which mark the four deepest minima of $\beta_{400}$}). As the scale $W$ decreases, we observe, for example in Fig.2(c) of Ref. \onlinecite{Chaos15}, that another more pronounced minimum of $\beta_{W}$ for $W\le 240$ appears before the 2005 Sumatra-Nias  $M$8.6 EQ (see Table \ref{tab2}){, but the brown arrows in Fig.2(c) of Ref. \onlinecite{Chaos15}  show that there are still two deep minima of $\beta_{240}$ which are either postseismic or precursory to smaller EQs}. For even smaller $W$, i.e., $W=100$ and $W=160$, two more minima become discernable before the 2010 Chile $M$8.8 EQ and the 2011 Tohoku $M$9.1 EQ.  Thus, a comparison of Fig.2 of Ref. \onlinecite{Chaos15} with Table \ref{tab2} reveals that before {\em all} the six $M\ge 8.5$ EQs minima of the variability $\beta$ of the order parameter $\kappa_1$ of global seismicity are observed. {Moreover, an inspection of the four deepest minima indicated by the arrows in Fig.2 \onlinecite{Chaos15}  shows that as $W$ decreases these deepest minima concentrate before the strongest EQs.} 

In order to clearly identify these minima, we focus on the smaller scales, i.e., $\beta_{100}$ and $\beta_{160}$, which, though more complex than those of larger scales, exhibit {their deeper} minima before all the $M\ge 8.5$ EQs. 
The situation then is similar to that in Japan where two scales $W=200$ and $W'=300$ have been used.\cite{PNAS13} Interestingly, the seismic rate in the Japanese catalog was\cite{PNAS13} 148 EQs/month leading to the fact that $W=200$ and $W'=300$  correspond to the average number of EQs in 1.35 and 2.03 months, respectively. The latter values do not differ much from 1.25 and 2.0 months calculated for $\beta_{100}$ and $\beta_{160}$ in the global CMT catalog (see Sec. \ref{sec3}), 
{thus supporting our hypothesis that the lead time of SES is a characteristic time scale for the solid Earth crust}. The selection of
 two scales $W$ and $W'$ is made because upon considering an almost {\em simultaneous} appearance of their local 
 minima  and examine if their ratio  $\min (\beta_{W'})/\min (\beta_{W})$  lies in the region around unity\cite{PNAS13,JGR14} a  variability minimum can be {unambiguously} identified.\cite{PNAS13,JGR14,PNAS15} This reflects the premise\cite{PNAS13} that there exists a {\em single }(critical) process of consecutive EQs -charactererized by the  ratio  $\min (\beta_{W'})/\min (\beta_{W})$- that leads to the variability minimum. After this variability minimum, the system tries to retain its equilibrium (cf. the curves in Fig.\ref{f2a} do not exhibit any significant trend although they consist of observables, 
 i.e., $\beta_W$'s, based 
 on a few months' data for an almost 39 years period) with a preseismic process  which eventually culminates with the occurrence of the strong EQ.  Here, 
in order to identify a local minimum of either $\beta_{100}$ or $\beta_{160}$ we require that this should be a minimum for at least its 15 previous and 15 future values. 
Thus, such a local minimum is at least a local minimum  for almost two weeks on average. Moreover, $\beta_{100}$ and $\beta_{160}$ should exhibit simultaneous minima. To secure this condition\cite{JGR14} 
we require at least 90\% of the EQs that are included in the calculation of the local $\beta_{100}$ minimum are also included in the calculation of the local $\beta_{160}$ minimum. Once simultaneous local $\beta_{100}$ and $\beta_{160}$ minima are observed, we examine whether their ratio $\min ( \beta_{160})/ \min (\beta_{100} )$ is close to unity as in the case of Japan.\cite{PNAS13,JGR14} When this occurs, we identify the so called variability  minimum. {Selecting the range
{$(r_1=)1.05 < \min ( \beta_{160})/ \min (\beta_{100} ) < 1.15(=r_2) $}, which was determined on the basis of the minima observed before the $M\ge 8.5$ EQs, we obtain the nine variability minima shown in Table \ref{tab1}.
For their determination, any value of $\beta_{100}$ below {$(\beta_0=)$}0.285 (which is the shallowest minimum $\beta_{100}$ value observed before an $M\ge 8.5$ EQ, see Table \ref{tab1})  was tested in order to examine whether it constitutes a local $\beta_{100}$ minimum. When such a local minimum was found, we investigated whether a value of $\beta_{160}$ that includes at least 90 EQs of the local $\beta_{100}$ minimum was a local $\beta_{160}$ minimum. If such a simultaneous 
local $\beta_{160}$ minimum was found, we examined if the ratio $\min ( \beta_{160})/ \min (\beta_{100} )$ lied  in the aforementioned range. If this was  the case then we have a variability minimum and the corresponding $\min ( \beta_{100})$ and $\min (\beta_{160} )$ values are indicated with the red and green circles in Fig.\ref{f2a}, respectively. }

\section{Discussion} \label{sec5}

An inspection of Table \ref{tab1} and Fig.\ref{f2a} shows that the variability minima in the global CMT catalog appear nine months at the most before 
the occurrence of magnitude class 9 EQs. Such a lead time is approximately three times larger than the corresponding lead time observed\cite{PNAS13,JGR14} in Japan for EQs of $M\ge 7.8$. This indicates that the preparation stage in global scale is presumably of larger duration than that in regional scale.

Moreover, a comparison of the present results with those obtained\cite{PNAS13,JGR14} from the regional study of Japan before the strongest EQ, i.e., the 2011 Tohoku $M$9.1 EQ,  leads to a compatible explanation:  
The variability minimum observed for this EQ in the global CMT catalog was on  29-30 November 2010. On the same 
date (see Table 1 of Ref. \onlinecite{PNAS13} where local time is used) a minimum was also observed in the regional Japanese catalog
that was associated\cite{PNAS13,JGR14} with the
2010 Near Chichi-jima EQ that occurred 22 days later. The deepest variability minimum for the 27 year period studied\cite{PNAS13,JGR14} in Japan was 
observed on 5 January 2011 being almost simultaneous\cite{SKO14} with the anomalous geomagnetic diurnal variations recorded\cite{XU013} at Esashi station at about 135km away from the Tohoku epicenter. Thus one may assume\cite{SKO14} that on the date of the deepest variability minimum in Japan an SES activity was {recorded} (whose magnetic field component\cite{SAR02,VAR03} was recorded at Esashi).  Then, the natural time analysis of seismicity\cite{SAR08,NEWBK} around the 
Tohoku epicenter leads\cite{SKO14} to the conc{l}usion that the criticality condition\cite{PNAS} $\kappa_1=0.07$ was fulfilled 31 hours before the occurrence of the 2011 Tohoku $M$9.1 EQ.
These results strengthen the view that the (regional) variability minimum is observed simultaneously with the emission of an SES activity as found in Ref. \onlinecite{TECTO12}. If we now take into account the observation\cite{ORI14} that the level and temperature of confined groundwater in north eastern Honshu (Japan's largest island) showed 
anomalous changes three months before the 2011 Tohoku $M$9.1 EQ (starting approximately on 10 and 15 December 2010, respectively, see their Fig.1) and such changes are precursory to large subduction EQs, we may have the view that the variability minimum in the global CMT catalog indicated the beginning of a preparatory process in a global scale{, presumably related with long-range spatial correlations between EQ activity patterns  (cf. such patterns\cite{TEN12} have been found\cite{TECTO12} very important for the appearrence of regional variability minima),   that was} followed within 10-15 days by the aforementioned changes in confined groundwater which was later followed by the simultaneous appearance of the 
regional variability minimum and the SES emission, and finally by the 2011 Tohoku $M$9.1 EQ which is the largest EQ both regionally and globally. 
The fact that the minimum  before Tohoku EQ was not found to be  the deepest when using CMT, might be understood in the following context:
In global scale, the Tohoku EQ magnitude does not differ much from the other $M\ge 8.5$ EQs that occurred in other regions of the world, but 
it does so in Japan where the second larger EQ was the 1994 East-Off Hokkaido $M$8.3 EQ and the relation between the value of $\min(\beta_{100})$ and the magnitude of the impending EQ is not very sharp (for example $\min(\beta_{100})$ before the 2014 Sumatra-Andaman $M$9.0 EQ is shallower than that before the 2005 Sumatra-Nias $M$8.6EQ and a similar picture holds also for Japan, see Table 1 of Ref. \onlinecite{PNAS13}, where the aforementioned 1994 East-Off Hokkaido $M$8.3 EQ exhibits the shallowest of the variability minima identified there). It is also noteworthy, that if the study of the global CMT catalog was made until the occurrence of the the 2011 Tohoku $M$9.1 EQ,  then an inspection of Table \ref{tab1} shows that one could have selected instead of 0.285 the value 0.277 (that preceded the 2007 Sumatra, Indonesia $M$8.5 EQ) to search for $\beta_{100}$ minima and in that case all EQs with $M\ge 8.5$ would have been preceded by variability minima (including the 2011 Tohoku $M$9.1 EQ) without any false alarm. By the same token, if such a selection was made until 1 October 2014, only the 2012 Indian Ocean $M$8.6 would have been missed leading to a hit rate of 80\% with no false alarms.

We now turn to the statistical significance of the variability minima in the global CMT catalog reported in Table \ref{tab1}. As it was done\cite{Sarlis2016} in the case of the Japanese catalog, we follow the 
Receiver Operating Characteristics (ROC) method.\cite{FAW06} ROC is a technique to 
depict the quality of binary predictions. It is a plot of the hit rate (or True positive rate)
versus the false alarm rate (or False positive rate), as a function of the total rate of
alarms, which is tuned by a threshold in the predictor.  The hit rate is the ratio 
of the cases for which the alarm was on and a significant event occurred over the total number of significant events. The false alarm rate is the ratio of the cases for which the alarm was on and no significant event occurred over the total number of non-significant events. 
Only if the hit rate exceeds the false alarm rate, a predictor
is useful (for example, the ROC analysis has been recently 
used\cite{TEL14} to disciminate between seismograms of tsunamigenic and 
non-tsunamigenic EQs).
Random predictions generate on average equal hit and false alarm rate  (thus, falling on the blue diagonal in 
Fig. 4 of Ref. \onlinecite{Chaos15}),
and the corresponding ROC curves exhibit fluctuations which depend on the 
positive $P$ cases (i.e., the number of significant events) and the negative $Q$ cases (i.e., the number of non-significant events) to be predicted.  The statistical significance 
of an ROC curve depends\cite{MAS02} on the area $A$ under the curve in the ROC plane. It has been shown\cite{MAS02} that 
$A=1-{U}/(PQ),$
where $U$ follows the Mann-Whitney U-statistics.\cite{MAN47} Recently, a visualization scheme 
for the statistical significance of ROC curves has been proposed.\cite{SARCHRIS14} It is based on k-ellipses 
which are the envelopes of the confidence ellipses -cf. a point lies outside a confidence ellipse with probability {$\exp (-{\rm k}/2)$-} obtained 
when using a random predictor and vary the prediction threshold. These k-ellipses cover the whole ROC plane 
and upon using their $A$ we can 
have a measure\cite{SARCHRIS14} of the probability $p$ to obtain 
by chance (i.e., using a random predictor) an ROC curve passing through each point of the ROC plane.
In order to apply the ROC method, we divide the whole period covering 465 months into 51
nine-month periods (i.e., $P+Q=51$) out of which only 6 included significant events $(P=6)$, i.e.,
EQs with $M\ge 8.5$, see Table \ref{tab1}.  Hence, the hit rate is 100\%. On the other hand, the 3 minima which were followed by smaller 
EQs (see the fourth, eighth and ninth line of Table \ref{tab2}) may be considered as false alarms giving 
rise to a false alarm rate of $3/45\approx6.67\%$. 
By using the FORTRAN code {\tt VISROC.f} provided by Ref. \onlinecite{SARCHRIS14} we obtain: 
(a) the ROC diagram of Fig. 4 of Ref. \onlinecite{Chaos15} in which we depict by the red circle the operation point 
that corresponds to the results presented in Table \ref{tab1} and (b) the probability $p$ to 
obtain this point by chance based on k-ellipses which results in $p=0.0047\%$. Interestingly, this $p$-value is of the order $10^{-5}$,  thus being 
similar with that found\cite{Sarlis2016} for the variability minima in the regional study\cite{PNAS13} of Japan. {Moreover, when we vary any of the three parameters $(\beta_0,r_1,r_2)$ that determine the operation point in order to obtain an ROC curve the resulting $p$-values estimated on the basis of the corresponding $A$-values lead to $p$-values ranging from $1.3 \times 10^{-4}$ to $6.5 \times 10^{-5}$, see the red, blue and cyan lines in Fig.4 of Ref. \onlinecite{Chaos15}. Here, it is worthwhile to  comment also on fact that the average value of the lower and the upper limit of the ratio of $\beta$ minima, $(r_1+r_2)/2(=1.1)$, almost coincides with the value of the same ratio observed before  the 2011 Tohoku $M$9.1 EQ which is the strongest EQ. This property -which is also observed\cite{PNAS13,JGR14} in the case of Japan- certainly points to the existence of a single (critical) process which is best monitored for the strongest EQ. It is also of interest to investigate whether such a prediction scheme could be extended to smaller target magnitude thresholds. Such an attempt is made in the Appendix. }

{In general, we have to stress the following: First, from the present analysis we may infer that a major EQ is likely approaching, but we cannot gain information on the likely epicenter location. This is clearly something to further consider in future work, probably by employing a procedure similar to that used\cite{PNAS15,HUANGPNAS15} to determine the future EQ epicentral location in the regional study of Japan. Similarly, the exact timing of the EQ cannot be predicted (just the average lead time of the obtained dynamical signatures is known from the present analysis, which varies from case to case). Second, the parameters $(\beta_0,r_1,r_2)$  used for identifying precursory signatures have been optimized a posteriori based on a few major events (see also the Appendix). Detailed long-term observations will be required to further constrain these parameters (based on a larger number of strong events) and assess whether the observed signatures are sufficiently stable  among different regions and EQ types.} 

Finally, we discuss the following two important points emerged in  the present study.
First, the scales $W$ that gave rise to the minima identified both globally and regionally correspond to the number of EQs that on average occur within one to two months. Such time scales have been first identified\cite{VAR91,VAR93}  as lead times for SES activities. Second, the average lead time of the variability minima in the global CMT catalog is five and a half months which strikingly agrees with
the maximum lead time observed for SES activities.\cite{SPRINGER} As for the regional study of Japan,\cite{PNAS13,JGR14} the corresponding average lead time of the minima is 2.25 months that
compares favorably with the average lead time of SES activities.\cite{SPRINGER}

\section{Conclusions}\label{sec6}
The Global Moment Tensor Catalog in the period 1 January 1976 to 1 October 2014 was investigated here for the existence of precursory variability $\beta$ minima of the order parameter $\kappa_1$ of seismicity in natural time as it was previously found\cite{PNAS13} for the strong EQs in Japan. 
If we consider appropriate natural time scales $W$, corresponding to the number of events that on average occur within one to two months, we were able to identify minima of $\beta_{100}$ and $\beta_{160}$ that precede all the six magnitude class 9 EQs ($M\ge 8.5)$. Their lead time varies from 1 to 9 months before the main shock and using their properties we separated them from all other similar local minima of  $\beta_{100}$ and $\beta_{160}$ except of 3 cases in 2008, 2012 and late 2013. 
The probability to obtain such a selection of precursory  phenomena by chance  is of the order of 10$^{-5}$.  The present results
strengthen the view that the solid Earth crust behaves as a single complex system. 

{
\appendix
\section{Possible extension to smaller target magnitude thresholds}
In the main text we presented a prediction scheme focused on the occurrence time of magnitude class 9 EQs, here we attempt an extension of this method to smaller target magnitude thresoholds. As a first step, we focus on the prediction of the occurrence time of the single $M=8.4$ EQ in global CMT which is the 2001 Peru $M$8.4 EQ, see Fig.5(b) of Ref. \onlinecite{Chaos15}. This EQ exhibits a $\min(\beta_{100})$ value equal to 0.352 (see the inset of Fig.5(b) of Ref. \onlinecite{Chaos15}) whereas the corresponding ratio  ${\min ( \beta_{160})}/{\min (\beta_{100})}\approx1.09$ lies within the narrow range $[1.05,1.15]$ used for the prediction of $M \geq 8.5$ EQs (see Table \ref{tab1}) when using $\beta_0=0.285$. Increasing $\beta_0$ to 0.353 leads to 17 additional minima followed by smaller EQs (false alarms) giving rise to the uppermost point of the ROC diagram depicted in Fig.6 of Ref. \onlinecite{Chaos15} (cf. a selection of $r_1=1.060$ and $r_2=1.135$, which is also compatible with the results of Table \ref{tab1} but has not been selected based on the fact that one should also allow for a plausible experimental error of 0.01 in the ${\min ( \beta_{160})}/{\min (\beta_{100})}$ ratio,  leads in this case to only 12 false alarms). An inspection of Fig.6 of Ref. \onlinecite{Chaos15} also reveals that such a scheme remains statistically significant with a $p$-value $1.4\times 10^{-4}$, comparable to the one presented in the main text. }

{When attempting to lower the target magnitude threshold below 8.4, e.g., to 8.3 (thus attempting to predict 12 EQs), a problem arises since the 1977 Sumbawa (Indonesia) $M$8.3 EQ occurred almost one and a half year after the beginning of the CMT catalog and the calculated $\beta_{160}$ values are available for only six months before this EQ, see Fig.5(a) of Ref. \onlinecite{Chaos15}. This constrains  $\min(\beta_{100})$ to $\min(\beta_{100})=$0.484 (see the dashed square in Fig.5(a) of Ref. \onlinecite{Chaos15}) with a ratio ${\min ( \beta_{160})}/{\min (\beta_{100})}\approx1.04$,  whereas a smaller value $\min(\beta_{100})$=0.459 observed early January 1997  cannot be used since there are no $\beta_{160}$ values to be paired with it (cf. there exist only 108 EQs in the CMT catalog for 1976). We mention that although the value of the ratio is close to the range  $[1.05,1.15]$  used for the prediction of $M \geq 8.5$ EQs the value of $\min(\beta_{100})=$0.484 is very high (cf. the mean value and the standard deviation of $\beta_{100}$ for the whole period studied is $\mu=$0.593 and $\sigma=$0.220, respectively). The use of $(\beta_0,r_1,r_2)=(0.484,1.04,1.15)$ in this case results in very frequent alarms amounting to $\tau$=47\% of the total time (when assuming that the alarm is stopped upon the occurrence of an $M\geq 8.3$ EQ) and a hit rate of 83.3\%(=10/12). However, by increasing only $r_2$ to 1.19 -thus selecting $(\beta_0,r_1,r_2)=(0.353,1.05,1.19)$- one can obtain  the same hit rate with $\tau=$32\%, while when  
$(\beta_0,r_1,r_2)=(0.353,1.05,1.32)$ a hit rate of 91.7\%(=11/12) is achieved with $\tau=$37\%. 
These results indicate that the prediction scheme becomes less efficient upon focusing on lower target magnitude thresholds. This is not incompatible with the view that EQs  with $M<8.5$ may or may not lead the whole Earth crust system to criticality.}

\begin{figure*}
\includegraphics[scale=0.65]{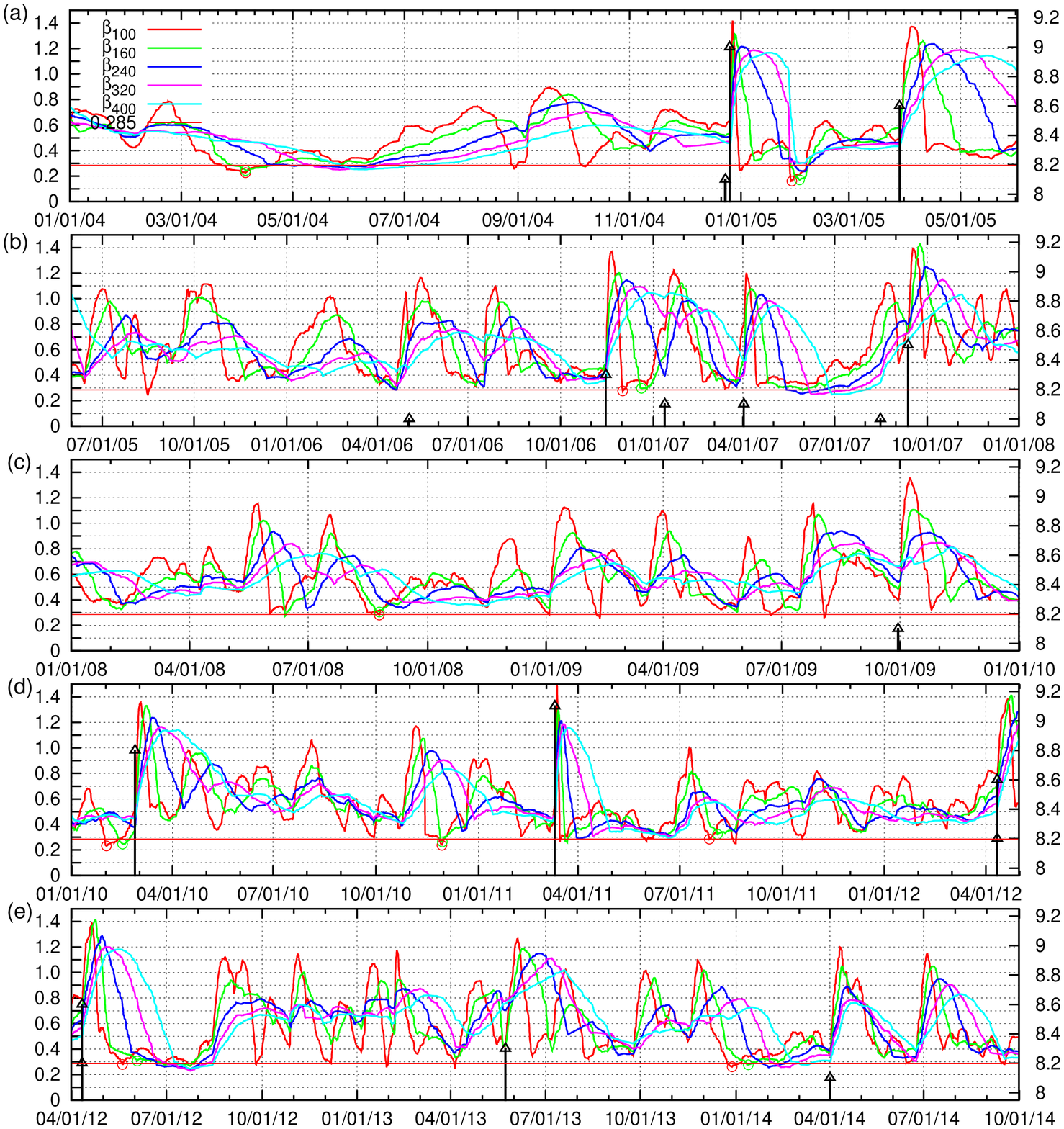}
\caption{(color online) The variabilities(left scale)  $\beta_{100}$ (red), $\beta_{160}$ (green), $\beta_{240}$ (blue), $\beta_{320}$ (magenta), and $\beta_{400}$ (cyan)  versus conventional time during the period 1 January 2004 to 1 October 2014. The red and green open circles indicate the identified $\beta_{100}$ and $\beta_{160}$ minima in each case.  The horizontal red line corresponds to $\beta_{0}=0.285$ which is the shallowest $\beta_{100}$ minimum observed before an $M\ge 8.5$ EQ in the  CMT catalog (it was observed before the 2012 Indian Ocean $M$8.6 EQ, see panel (d) and Table \ref{tab1}). 
The EQs with $M \geq 8.0$ (right scale) are also shown with the vertical lines ending at black triangles.}\label{f2a}
\end{figure*}

{If we further decrease the target magnitude threshold down to 8.0, the number of EQs to be predicted within the almost 39 years period of our study increases to 30. We cannot any more employ a method similar to the one presented above in which the alarm was stopped upon the occurrence of an $M\geq 8.3$ EQ, because for example three days before the 2004 Sumatra-Andaman $M$9.0  EQ an $M$8.1 EQ took place, see Fig.\ref{f2a}(a). 
For this reason, we considered the case that the alarm lasts exactly nine months after the observation of $\min(\beta_{160})$ indepent of whether a strong EQ takes place or not. In this case, the aforementioned selection of $(\beta_0,r_1,r_2)=(0.353,1.05,1.19)$ leads to a hit rate of 66.7\% with $\tau=38\%$. Alternatively, by decreasing $r_1$ to 1.04 -and hence using $(\beta_0,r_1,r_2)=(0.353,1.04,1.19)$- we can obtain a hit rate of 73.3\% with $\tau=41$\% (see Fig.5 of Ref. \onlinecite{Chaos15}). Interestingly, in the latter case, $(r_1+r_2)/2$ lies very close to the value of the ratio ${\min ( \beta_{160})}/{\min (\beta_{100})}$ observed before the 2011 Tohoku $M$9.1 EQ, see Table \ref{tab1}. An observation of Figs. 5(a) and 5(b) of Ref. \onlinecite{Chaos15} reveals that during the period from 1976 to 2002, when the total number of EQs with $M\geq 8.0$ was only thirteen, the total alarm time amounted to $\tau=$22\% with a hit rate 7/13$\approx 54$\%. However, Fig.5(c) of Ref. \onlinecite{Chaos15},  that covers the period 2002 to 2014 comprising of seventeen $M\geq 8.0$ EQs, reflects that the corresponding quantities are $\tau=$75\% and 15/17$\approx 88$\%. These numbers reveal that even for EQs with $M\geq 8.0$ such a scheme performs fairly well in both quiet and seismic active periods since the obtained hit rates are above 50\% and larger than their corresponding $\tau$ values.  }


\clearpage

\begin{table}
\caption{All EQs with $M\geq 8.5$ during the period 1 January 1976 to 1 October 2014 together with the outcome of the binary 
prediction scheme discussed in Section \ref{sec5}.}
\label{tab2}       
\begin{tabular}{cccccc}
\hline
EQ date  &  area   &  Lat. ($^o$N)   &  Long. ($^o$E)   &  $M_w$    &  outcome \\
\hline
2004-12-26    &  Sumatra-Andaman   &  3.30    &   95.78    &   9.0    &  TP\\
2005-03-28    &  Sumatra-Nias    &  2.09    &   97.11    &   8.6    &  TP\\
2007-09-12    &  Sumatra, Indonesia   &  -4.44    &   101.37    &   8.5    &  TP\\
2010-02-27    &  Chile   &  -35.85    &   -72.71    &   8.8    &  TP\\
2011-03-11    &  Tohoku, Japan   &  38.32    &   142.37    &   9.1    &  TP\\
2012-04-11    &  Indian Ocean   &  2.33    &   93.06    &   8.6    &  TP\\
\hline
\end{tabular}
\end{table}


\begin{table}
\caption{The nine  variability minima that were found
during {the whole study period together with the strongest EQ that followed $\min( \beta_{160})$} within nine months ($\Delta t_{100}$ and $\Delta t_{160}$ denote the corresponding  lead times in months for $W=100$ and $W'=160$, respectively).}
\label{tab1}       
\begin{tabular}{cccccccc}
\hline
$\min ( \beta_{100})$  (date)    &  $\min ( \beta_{160})$ (date)           &  $\frac{\min ( \beta_{160})}{\min (\beta_{100})}$    &  EQ date    &  EQ  area    &  $M$   &  $\Delta t_{100}$     &  $\Delta t_{160}$   \\
\hline
  0.2270(2004-04-05)   &  0.2431(2004-04-05)   &  1.071     &   2004-12-26    &  Sumatra-Andaman          &   9.0    &    8.8    &    8.8     \\
  0.1605(2005-01-28)   &  0.1702(2005-02-02)   &  1.060     &   2005-03-28    &  Sumatra-Nias    &   8.6    &    2.0    &    1.8     \\
  0.2770(2006-12-02)   &  0.2971(2006-12-20)   &  1.073     &   2007-09-12    &  Sumatra, Indonesia    &   8.5    &    9.5    &    8.9     \\
  0.2789(2008-08-25)   &  0.3049(2008-08-25)   &  1.088     &   2009-01-03    &  Papua, Indonesia    &   7.7    &    4.4    &    4.4     \\
  0.2317(2010-02-01)   &  0.2462(2010-02-16)   &  1.063     &   2010-02-27    &  Chile   &   8.8    &    0.9    &    0.4     \\
  0.2368(2010-11-29)   &  0.2638(2010-11-30)   &  1.114     &   2011-03-11    &  Tohoku, Japan    &   9.1    &    3.4    &    3.4     \\
  0.2849(2011-07-27)   &  0.3232(2011-08-04)   &  1.134     &   2012-04-11    &  Indian Ocean        &   8.6    &    8.6    &    8.4     \\
  0.2789(2012-05-20)   &  0.3053(2012-06-03)   &  1.095     &   2013-02-06    &  Solomon Islands          &   7.9    &    8.7    &    8.3     \\
  0.2612(2013-12-28)   &  0.2767(2014-01-13)   &  1.059     &   2014-04-01    &   Iquique, Chile    &   8.1    &    3.1    &    2.6     \\
\hline
\end{tabular}

\end{table}


\end{document}